\documentclass[twocolumn]{article}

\usepackage{graphicx}   
\usepackage[utf8]{inputenc}
\usepackage{amsmath}
\usepackage{tikz}
\usepackage{float}
\usetikzlibrary{arrows,intersections}
\usepackage{relsize}

\usepackage{graphicx,wrapfig,lipsum}
\usepackage{pgfplots}

\usepackage{tabularx}
\usepackage{amsfonts}
\usepackage{import}
\usepackage{cite}
\graphicspath{{}}

\pgfplotsset{width=10cm,compat=1.10}
\newcommand{\hspm}{\hspace{-2pt}}

\begin{document}

\title{Estimating load points of a motor-pump system using pressure and inverter drive data} 

\author{Yashar Kouhi,
    Jens Müller,
    Sebastian Leonow, and Martin Mönnigmann
\thanks{The authors are with the Chair of Automatic Control and Systems Theory, Ruhr-Universität Bochum, Germany. Address:
 Universitätsstraße 150, 44801 Bochum. E-mail: \{yashar.kouhi, jens.mueller-r55, sebastian.leonow, martin.moennigmann\}@rub.de.}}%

\maketitle

\begin{abstract}
 We propose a novel method for the estimation of rotor position, speed, and torque of a motor-pump system consisting of a progressive cavity pump (PCP) driven by an induction motor which operates under V/f open-loop control.
We compute the speed and rotor position of the PCP by applying a phase locked loop (PLL) to the pressure signal at the 
pressure side of the pump. 
An extended Kalman filter is used to estimate the torque of the PCP based on the speed, effective value of the stator current of the induction motor and a nonlinear motor model. 
Furthermore, we derive a tractable condition under which the   convergence of  the observer is guaranteed. We use a laboratory experiment to verify our results.
\end{abstract}


\maketitle

\section{Introduction}
Progressive cavity pumps (PCPs) have a broad range of applications in the food, cosmetics, and petroleum industries, 
where they are used for fluids with high viscosity and abrasive media in particular \cite{witt2012}. 
Due to the practical importance of PCPs, model-based methods for flow estimation~\cite{Andrade2010,PALADINO2011178,Zeng18},
modeling of torque in the presence of negligible viscosity and frition effects \cite{witt2012,ZHOU201371}, 
and condition monitoring and wear detection~\cite{Mueller2020} have attracted sustained attention. 
All these methods require to know the pump speed, torque, or rotor position, which are typically measured by expensive sensors such as absolute encoders or torque transmitters. It is of obvious interest to replace these sensors by estimation methods whenever possible. 

A PCP is usually driven by a variable frequency inverter and an induction motor. 
Because the pump and motor  are connected by a common shaft, they share the same  values of the rotor position, speed, and torque. 
Speed, torque, and rotor position estimation of induction motors is a well studied topic in the context of  electric drives. 
The most prominent approaches for speed estimation of induction motors use the equivalent circuit of an induction motor for finding the slip between the synchronous speed and the rotor speed in steady state \cite{sens2007}, 
directly compute the speed through the estimated rotor flux \cite{KanTaa93},
or are based on model reference adaptive systems \cite{kub93} or extended Kalman filters \cite{shietal02, Kim94}. 
Once the speed is estimated, the torque estimation can be accomplished from the steady state model \cite{sens2007} or the dynamic model of the motor \cite{kub93}. 
Methods based on the anisotropy in the motor air gap have also been proposed 
for estimating speed and rotor position of induction motors. 
These techniques use the spatial harmonics of the rotor position on the stator current \cite{HaSul99,Holz02}. 
However, some special engineering modifications of the rotor or stator slots are usually required for anisotropy-based methods to work properly for induction motors \cite{Holtz98}. 

It is not straight forward to use any of the existing methods in hydraulic processes. 
All of these methods require a high-resolution current measurement of at least one phase of the motor, but current sensors are often not available in hydraulic applications due to their additional cost. 
Anisotropic effects are not significant in standard induction motors, which renders rotor position estimation based on anisotropic methods impractical for the standard motors used in hydraulic applications.

In this contribution, we introduce a new method for estimating rotor position, speed, and torque   of a PCP driven by an induction motor under voltage/frequency (V/f) open-loop control. 
We treat the motor and pump as two coupled subsystems and use signals from both subsystems for the purpose of estimation. More specifically, our algorithm requires the pressure signal at the pressure side of the PCP provided by a pressure sensor, 
and the effective value of the stator current of the motor measured by the variable frequency inverter (VFI). %
Note that the approach requires a hardware pressure sensor, which, however, is considerably less expensive than, e.g., an absolute encoder or a torque transmitter. Moreover, pressure sensors are already available in many hydraulic processes. 
We use a phase locked loop (PLL) to compute the speed and rotor position from the second harmonic of the pressure signal. This step exploits the periodicity of the pressure signal of the PCP and its dependence on the rotor position. 
The estimated speed can be combined with the effective value of the motor current for an observer-based estimation of the torque. 

We introduce the parameters and signals used of the induction motor and the PCP in the remainder of this section. 
In Section\,2, we present  our algorithm for the estimation of the rotor position, speed, and torque of the motor-pump system. In Section\,3 we introduce the process control system and verify our algorithm by implementing it on a 
test setup. %
We give a brief conclusion in Section\,4.

\noindent\textbf{Nomenclature}\\[.1cm]
\textit{Motor variables:}\\
$R_s$ and $R_r$: stator and rotor resistances \\
$L_s$ and $L_r$: stator and rotor inductances\\
$L_m$: mutual inductance\\
$\sigma= 1-\tfrac{L_m^2}{L_s L_r} $: leakage coefficient\\
$T_s= \frac{L_s}{R_s}$ and $T_r= \frac{L_r}{R_r}$: stator and rotor time constants\\
$J$: moment of inertia\\
$F$: friction coefficient\\
$z_p$: number of pole pairs\\
$\omega_s$, $\omega_e$ and $\omega_m$: synchronous, electrical, and mechanical rotational speeds\\
$s_m$ and $s_m \omega_s$: slip and slip rotational speed\\
$i_{sd}$ and $i_{sq}$: stator currents in $d$ and $q$ axes\\
$\psi_{rd}$ and $\psi_{rq}$: rotor magnetic fluxes in $d$ and $q$ axes\\
$u_{sd}$ and $u_{sq}$: stator voltages in $d$ and $q$ axes \\
$u_s = [u_{sd} ~ u_{sq}]^\top$: stator voltage vector \\
$i_{\text{eff}}$: effective value of stator current\\
$T_e$: electromagnetic torque\\
$T_L$: load torque\\[.2cm]
\textit{PCP and gearbox variables:}\\
$\theta_p$: pump rotor position\\
$\omega_p$: rotational speed \\
$n_p$: pump speed in revolutions per minute\\
$T_p$: pump torque\\
$p_D$: discharge pressure at the pressure side of the pump\\
$\overline{p_D}$: mean value of discharge pressure for one revolution of the rotor\\
$\gamma$ and $\eta$: transmission ratio and efficiency of a gearbox
\section{ Soft sensor algorithms for the motor-pump system}
\begin{figure}[t]
\includegraphics[width=8.4cm]{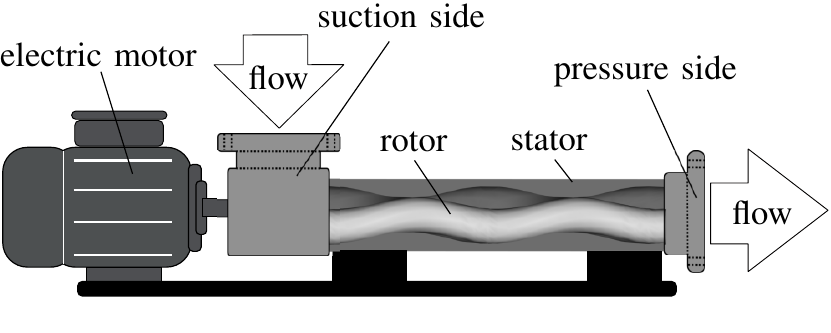}
\caption{Sketch of the motor-pump system. } \label{fig:pumpmotor}
 \end{figure}

We consider single-stage single-lobe PCPs driven by induction motors and VFIs as shown in Fig.\,\ref{fig:pumpmotor}. We assume that the discharge pressure at the 
pressure side of the pump ($p_D$) is measured by a pressure sensor and the effective value of the motor current $i_{\text{eff}}$ is measured by the VFI. We further assume that the motor rotates only in one direction and operates under V/f open-loop control. The  V/f control keeps 
the ratio of the magnitude of voltage vector to frequency, i.e., $\|u_s\|/\omega_s$, constant. The  magnetic flux in the motor air gap will then be almost constant and therefore the motor can function at all operating points in the entire constant torque region \cite{sens2007}. We emphasize that under V/f control the motor speed is not exactly equal to the synchronous speed, and a slip between these two variables results. The amount of slip 
depends on the pump torque.  

In this section, we assume for simplicity that the motor and pump have the same speed, i.e., $\omega_p = \omega_m$, $T_L = T_p$, and that the motor has two pole pairs $z_p=2$, which implies $\omega_e = 2\omega_m$. Otherwise, if the PCP and the motor with pole pairs $z_p$ are connected by a gearbox with the transmission ratio $\nu$ and efficiency $\eta$, the substitutions $\omega_p = \omega_m /\nu$, $T_p = \eta \nu T_L$, and $\omega_e = z_p \omega_m$ must be made. 

We present our estimation algorithm for the speed and rotor position of the PCP based on the PLL technique in Section\,\ref{sec:pll}. Then, we address the torque estimation problem in Section\,\ref{sec:torque}. 
 \subsection{ PLL design for the estimation of the rotor position and speed of the PCP from the pressure signal} \label{sec:pll}
\begin{figure}[t] 
\hspace*{.5cm} \includegraphics[width=6.4cm]{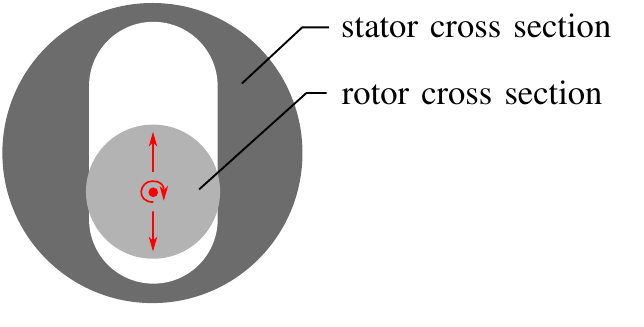}
\caption{Cross section of the  rotor inside the stator in a PCP.}\label{fig:corsssec}
 \end{figure}

\begin{figure}[t] 
  \centering
\resizebox{80mm}{!}{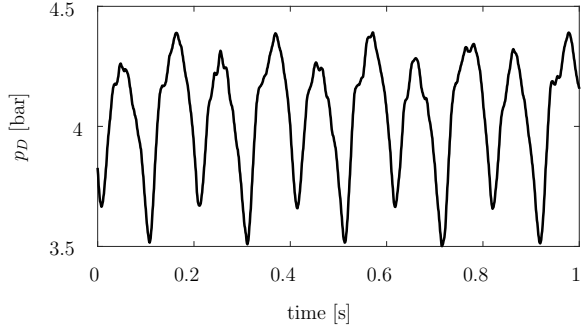}
 \caption{ Discharge pressure at the pressure side $p_D$ of a PCP with $\omega_p= 30.89$\,rad/s.} \label{fig:Pres_sig}
 \end{figure}

\begin{figure}[t] 
 \centering
 \resizebox{80mm}{!}{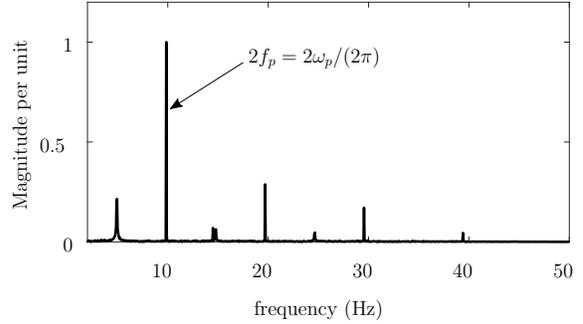}
 \caption{Fourier spectrum of the pressure signal shown in Fig\,\ref{fig:Pres_sig}.}\label{fig:FourierTran}
 \end{figure}

The conveying principle of PCPs relies on cavities that form between rotor and stator and transport fluid from the suction side to the pressure side. Two distinct motions of the rotor can be distinguished at the cross section of the 
pressure side of the PCP (see Fig.\,\ref{fig:corsssec}). Firstly, the rotor rotates around its center point, a rotation caused by the common motor-pump shaft. 
Secondly, for any full revolution of the motor-pump shaft, the cross section of the rotor oscillates in a translational motion along the slot hole of the stator at the pressure side of the pump. The frequency of both motions equals the rotational frequency of the motor-pump shaft $\omega_p$.
The rotor movement that results from the superposition of the two motions leads to a periodic opening and closing of a cavity between rotor and stator on the pressure side of the pump. The opening and closing cavity discharges the transported fluid and leads to a periodic fluctuation in the discharge pressure.
The pressure signal shows periodic extrema as a consequence. Figure\,\ref{fig:Pres_sig} shows a sample pressure signal as an illustration.  
Pressure minima occur whenever the rotor is located near one of the edges of the slot hole, which results in two such minima per revolution (see \cite{Belcher91} for a detailed discussion on the pressure pulsation). 
The signal contains higher-frequency harmonics, which is also evident from its Fourier frequency spectrum shown in Fig.\,\ref{fig:FourierTran}.

It is our 
aim to provide an estimate $\hat\theta_{p}$ of the rotor position (phase angle) from the measured discharge pressure $p_D$ and the synchronous motor speed $\omega_s$. We exploit the information contained in the pressure signal $p_D$ by using a PLL algorithm (see, e.g., \cite{Abram2002}) to determine the phase angle $\hat\theta_{p_D}$. Since the pressure $p_D$ oscillates with twice the frequency of the rotor, the relevant peak in the frequency spectrum (Fig. \ref{fig:FourierTran}) is located at $2f_p$. 
The rotor position $\theta_p$ and the phase angle of the discharge pressure $\theta_{p_D}$ are related by
\begin{equation}\label{eq:OffsetSubtraction}
\theta_p=\frac{1}{2}\theta_{p_D}-\theta_\text{off}\ ,
\end{equation}
where $\theta_\text{off}$ is an offset angle. We assume the offset angle to be almost constant for all operating points $(\omega_s,\ \overline{p_D})$, which is supported by the experimental results given in section \ref{sec:experiment}\footnote{The variation of $\theta_\text{off}$ with the increase/decrease of $\overline{p_D}$ (see Fig.\,\ref{fig:Druck2bar}) occurs in a range that is small compared to the domain $\theta_\text{off}\in [-90^\circ, 90^\circ]$ and does not affect the convergence of the proposed algorithm.}. 
The offset angle therefore needs to be determined only once by, e.g., determining the phase offset between $\theta_p$ and $\theta_{p_D}$ with an incremental encoder for the rotor position $\theta_{p}$.

The pressure signal is first filtered by a bandpass filter $G_F(s)$. As the frequency of the second harmonic of the pressure signal is twice the rotor speed and the motor has two pole pairs, we choose the bandpass frequency of the filter to be equal to $\omega_s$. As a convenient choice for the bandpass filter, we take a second order system in the form of
 
 \begin{align}
 G_F(s)& = \frac{2\pi B s}{s^2+2\pi B s +\omega_s^2},
\end{align}
where $B$ is the bandwidth of the filter. 
The parameter $B$ can be selected by considering  the  nominal slip rotational speed $s_{m,n}\omega_{s,n}$ of the motor which is indeed the maximal slip frequency in the entire constant torque region (\cite{Hughes13}, pp. 287). The filter phase reads
\begin{align} \label{eq:phase_filter}
 \theta_F& = \frac{\pi}{2} - \text{atan}\left(\frac{2\pi B {\omega}}{-{\omega}^2 +\omega_s^2} \right).
\end{align}

\begin{figure*}[t] 
 \hspace*{0.5cm}
 \centering{
 \includegraphics[scale=.92]{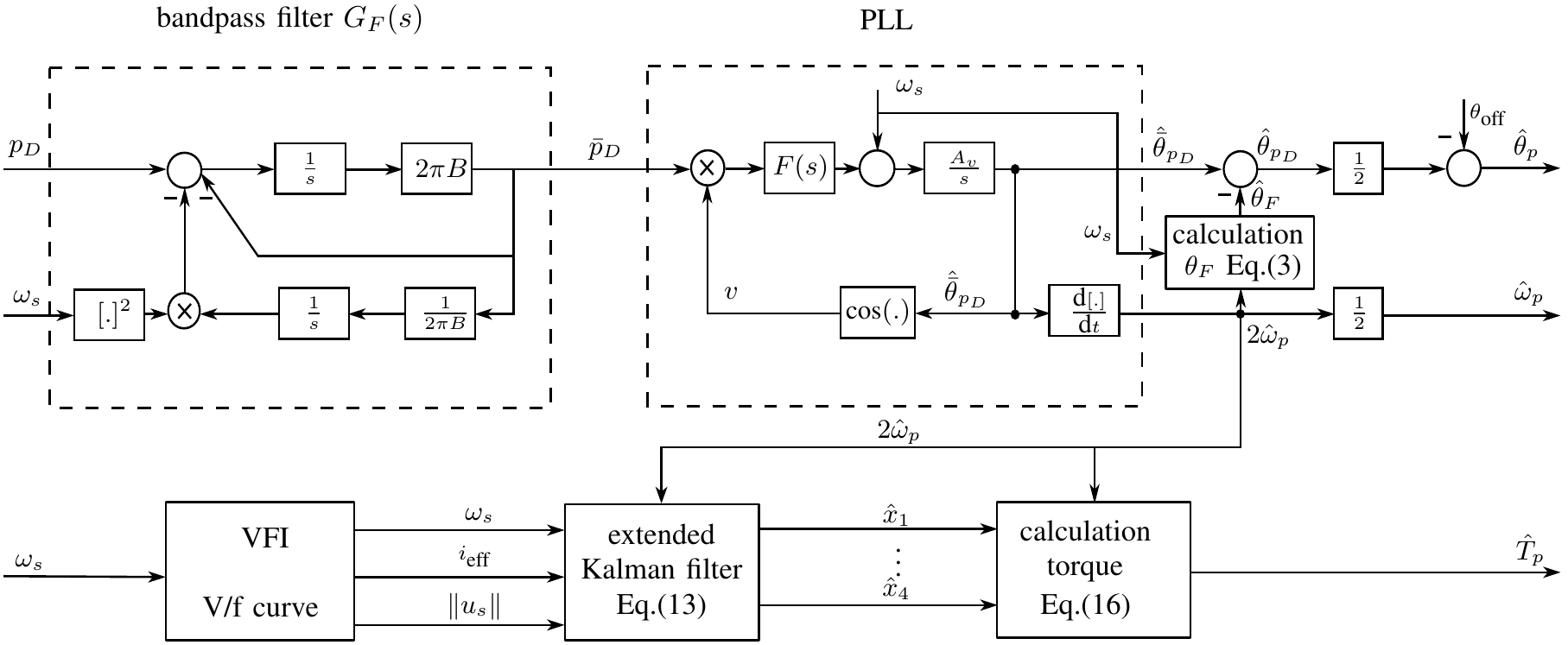}
 \caption{Block diagram for the estimation of the mechanical variables of a PCP under V/f control.}
 \label{fig:PLL22}
 }
 \end{figure*}

Block diagrams for the bandpass filter and PLL algorithm are given in the upper part of Fig.\,\ref{fig:PLL22}. 
 In the PLL block, $F(s)$ refers to a first order filter in the form of 
$
F(s) = 1/(Ts+1),
$
and $A_v$ is a constant. 
 The time constant $T$ and the gain $A_v$ can be selected in a manner that the damping factor for the closed-loop system of the PLL is equal to $\sqrt{2}/2$, resulting a 4\% overshoot to a step input \cite{Abram2002}. 
 
 %
 
Once the phase $\tfrac{1}{2} \hat \theta_{p_D}$ has been estimated, the offset phase $\theta_{\text {off}}$ must be subtracted according to 
 \eqref{eq:OffsetSubtraction} to determine $\hat{\theta}_p$ (upper right in Fig.\,\ref{fig:PLL22}). 
Furthermore, the derivative with respect to time must be calculated to obtain the estimate of the shaft frequency $\hat \omega_p$ (center right in Fig.\,\ref{fig:PLL22}).  
  \newcommand*{\MyPath}{/home/users/yak/Umbau_21_Projekt/Bericht_Umbau/Journal_Paper/Pics/BlockDiag}%
\subsection{Observer design for the estimation of the motor-pump torque} \label{sec:torque}
This section presents the extended Kalman filter for the estimation of the motor current and rotor flux vectors. The filter is based on an established model~(see, e.g., \cite{Quang:15}) for a three phase induction motor under V/f open-loop control, which we summarize as needed for the paper. 


In V/f control mode, the synchronous frequency $\omega_s$ is chosen by the operator to meet the requirements of the driven machine, i.e., the PCP in our case.  
 The magnitude of the stator voltage $\|u_s\|$ is known from the V/f table implemented by the VFI. 
 Therefore, it is convenient to use the dynamic model of an induction motor in the synchronous coordinate frame $d\hspm-\hspm q$ assuming that the $d$ axis is fixed to the stator voltage vector, since only the magnitude of the stator voltage is known. In other words, $u_{sd} = \|u_s \|$ and $u_{sq} = 0$ hold in this coordinate system. We define the state vector and the input of the induction motor as~\cite[pp.~75]{Quang:15}
\begin{align}
x(t) &= [x_1~ ~x_2~~x_3~~x_4]^\top:= [i_{sd} ~~i_{sq}~~\psi'_{rd}~~\psi'_{rq}]^\top,\\
 u(t)&= \|u_s\|,
\end{align}
where the states $\psi'_{rd}= (1/L_m) \psi_{rd}$ and $\psi'_{rq}= (1/L_m) \psi_{rq}$ are introduced to simplify the model. 
 
We assume that the effective value of the stator current $i_\text{eff}$ is measured by the VFI. 
Under V/f open-loop control, the current of each phase is almost sinusoidal (the effect of harmonics of the current due to space vector modulation is neglected). 
This implies the magnitude of the current in the stator voltage coordinate system equals $\sqrt 2\, i_{\text{eff}}$.
We consider  the square of $i_{\text{eff}}$ as the output measurement signal 
\begin{align} 
 y(t) = h(x)=  i_{\text{eff}}^2= \frac{1}{2}(i_{sq}^2 +i_{sd}^2)= \frac{1}{2}(x_1^2 +x_2^2).
\end{align} 

We use the well-established model for the field synchronous coordinate systems to describe the state space representation of the motor dynamics  \cite{Quang:15} (pp. 75). The input-output model of the induction motor in the voltage coordinate system is described by the 
 nonlinear time variant system 
\begin{align} \label{eq:statespace}
 \text{model:} \left\{\begin{array}{cl}
 \dot x(t) &= A(t) x(t) +B u(t), \vspace{.2cm}\\
 y(t) &=\frac{1}{2}(x_1^2+x_2^2), 
\end{array}\right.
\end{align}
where 
\begin{align} 
A(t)& = \begin{bmatrix}
   -\gamma_1 & \omega_s(t) & \gamma_2 & \gamma_3 \omega_e(t) \\
   -\omega_s(t) & -\gamma_1 &-\gamma_3 \omega_e(t) & \gamma_2 \\
   \frac{1}{T_r} & 0 & -\frac{1}{T_r} & \omega_s(t)-\omega_e(t)\\
   0 & \frac{1}{T_r} & -\omega_s(t)\hspm +\hspm \omega_e(t) & -\frac{1}{T_r}
   \end{bmatrix},\notag\\[.5cm] B & =
    \begin{bmatrix}
   \gamma_4 & 0& 0& 0  
   \end{bmatrix}^\top\hspm. 
 \end{align}
 The parameters $\gamma_1, \ldots, \gamma_4$ 
\begin{align}
 \gamma_1 &= \left(\frac{1}{\sigma T_s} + \frac{1-\sigma}{\sigma T_r} \right), \qquad & \gamma_2 = \frac{1-\sigma}{\sigma T_r},\\
 \gamma_3 & = \frac{1-\sigma}{\sigma},\qquad &\gamma_4 = \frac{1}{\sigma L_s}. 
\end{align}
depend on the motor parameters and therefore are constant. 
Because $\omega_s$ is known a-priori and because the electrical speed $\omega_e $ can be replaced by $\omega_e(t)= 2 \hat \omega_p(t)$, $A(t)$ is a known matrix. 
 
 The electromagnetic torque, given by $T_e = \frac{3}{2} z_p(1-\sigma) L_s \left(\psi'_{rd}i_{sq}-\psi'_{rq}i_{sd}\right)$, can be represented in terms of $x_1,\ldots,x_4$ as
 \begin{align} \label{eq:elc_mom}
 T_e = \frac{3}{2} z_p(1-\sigma) L_s \left(x_2x_3-x_1x_4\right).
 \end{align}
 
The continuous-time extended Kalman filter for the estimation of the states of \eqref{eq:statespace},
which reads
\begin{align}
 \text{estimator:}
 \left \{\begin{array}{cl}
 \dot {\hat {x}}(t) &= A(t) \hat x(t) + B u + K(t)\left(y(t) -\hat y(t)\right),\vspace{0.2cm}\\
  \hat y(t) &= \frac{1}{2} \left(\hat x_1^2+\hat x_2^2\right), 
\end{array}\right. \label{eq:observersyst}
\end{align}
where $\hat x$ and  $\hat y$ are the estimated states and output, respectively.  
The gain $K \in \mathbb {R}^4$ is obtained by solving the Riccati equation 
\begin{align}\label{eq:schazer}
 \dot P(t)\hspm &= A(t) P(t)\hspm +\hspm P(t)A(t)^\top\hspm\hspm-\hspm P(t) C(t)^\top R^{-1}C(t) P(t)\hspm+\hspm Q, \notag\\
 K(t) &= P(t)C(t)^\top R^{-1}, 
\end{align}
forward in time,
where $P(0)=\delta I$, $\delta\in\mathbb{R}$, $\delta > 0$ and $I$ is the identity matrix.  
The row vector $C$ is then calculated from
 \begin{align}
  C(t) &= \frac{\partial y}{\partial x}\Big|_{x = \hat x} = [\hat x_1~~\hat x_2~~ 0~~ 0].
 \end{align}
We assume $Q$ to be positive definite and $R= \kappa I$, $\kappa> 0$. 
Although the convergence of the extended Kalman filter is not always guaranteed \cite{Gelb201}, 
 the observer \eqref{eq:observersyst} asymptotically converges under the assumptions on $Q$ and $R$, 
if the estimated and measured currents share the same sign. This claim is proved in Appendix\,A, where we also show the sign criterion can be checked easily. 

%
With the estimated variable $\hat x$, the estimated torque $\hat T_e$ is calculated from \eqref{eq:elc_mom}.
Inserting $\omega_m= \hat \omega_p$ into the momentum equation $T_p = T_e-F \omega_m-J \dot { \omega}_m$, we derive the desired estimate
\begin{align} \label{eq:pcp_moment}
 \hat T_p &= \hat T_e- F\hat \omega_p-J \dot {\hat \omega}_p. 
\end{align}
 The torque estimation is illustrated in the lower part of Fig.\,\ref{fig:PLL22}.
\vspace{-.2cm}
\section{Experimental setup}
\label{sec:experiment}
\begin{figure}[t] 
\centering
\hspace*{0cm}
\includegraphics[scale= 1]{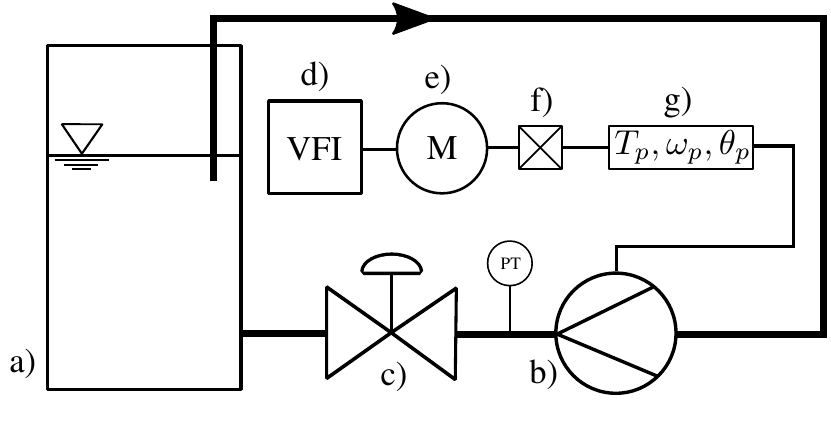}
\caption{Components of the pumping process test setup: a) container; b) PCP; c) control valve; d) VFI; e) induction motor; f) gearbox. Measurement devices: g) torque, speed, and rotor position transmitter; PT: pressure transmitter. 
}
\label{fig:testbench2}
\end{figure}

We apply the proposed algorithm to a test setup with the PCP model $10\text{-}6$L\footnote{The pump is manufactured by the company Seepex GmbH and is convenient for pumping media up to $6$\,bar.} driven by an induction motor\footnote{ The motor SK-112MH/4 is manufactured by the company Nord GmbH. The following nominal data are printed on the data plate of the motor: power $4$\,KW,  current $8$A in star connection, frequency $f_n = 50$\,Hz, effective voltage of one phase with respect to the star connection point $U_n=230$\,V,  speed $n_n = 1440$\,rpm, power factor $\text{cos}\, \varphi_n = 0.83$, and number of pole pairs $z_p=2$.}.
 The motor and pump are connected via a gearbox with the transmission ratio $\nu=2.94$ and the efficiency $\eta=0.96$.
 The motor parameters are 
\begin{align*}
 R_s&=R_r= 1.16\Omega,~ L_m = 0.2\, \text{H},\notag \\
 &\hspace{.5cm} L_s= L_r = 0.21\,\text{H},~ \sigma = 0.081.
\end{align*}
The friction coefficient of the shaft is taken from the motor manufacturer data sheet as $F =7.69\times 10^{-4}$.
The VFI is set to V/f open-loop mode. 
Figure\,\ref{fig:testbench2} shows a sketch of the experimental setup. 
The fluid pumped from the container\,a) passes through the PCP\,b) and the control valve\,c), and flows back to the container\,a). 
The control valve\,c) is used to adjust the discharge pressure $p_D$ and thus the operating point of the pump. For the verification of 
the results the speed, torque, and rotor position of the pump are also recorded separately by extra sensors. All data are gathered by the software TwinCAT\footnote{TwinCAT is an automation software from the company Beckhoff GmbH and is appropriate for real-time control and online data measurement of a process.} with a sample rate of $1$\,ms. 

\begin{figure}[t]
\hspace*{-.5cm}
 \resizebox{.54 \textwidth}{!}{\fontsize{14}{16} \selectfont 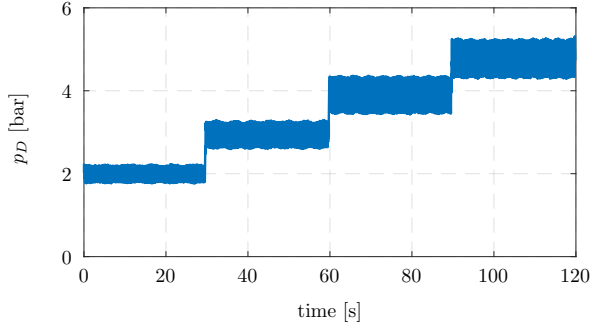}
\caption{Pressure signal of the PCP during the experiment with the variable speed and constant position of the valve c). }
\label{fig:pressuresignalexp}
 \end{figure}
\begin{figure*}[h]
\hspace*{-1cm}
 \centering{
 \resizebox{1.11 \textwidth}{!}{\fontsize{14}{16} \selectfont 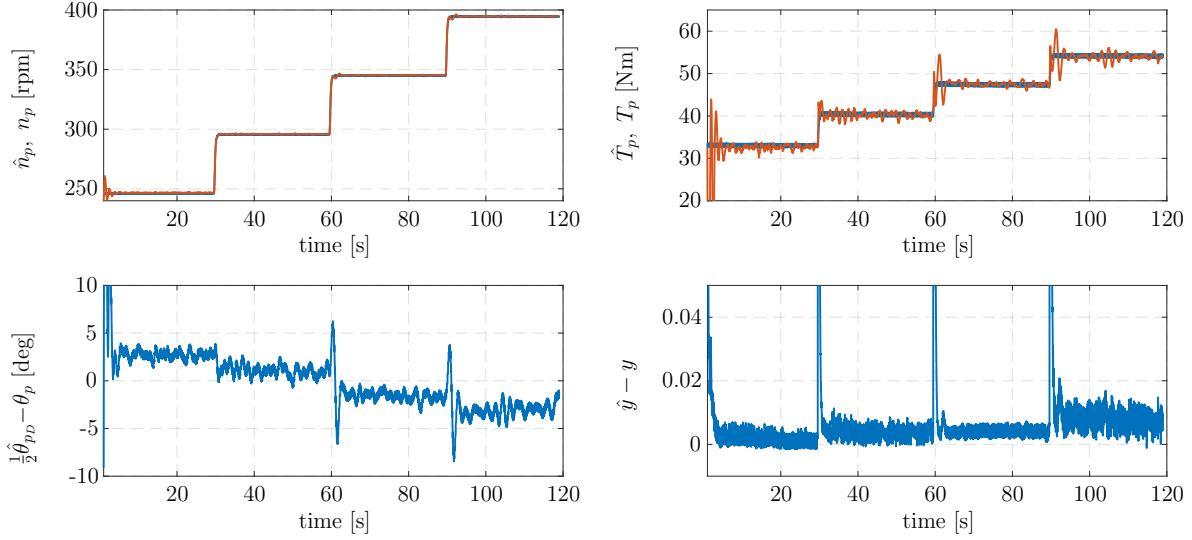}
 \caption{Upper diagrams: estimated (orange color) and measured signals (blue color), lower diagrams: error signals.}
\label{fig:Druck2bar}
 }
\end{figure*} 
In our experiment, starting from the motor synchronous frequency $ \omega_s/(2\pi)=25$\,Hz, we increment the synchronous frequency with a step size of $5$\,Hz every $30$\,s, while keeping the position of the control valve\,c) constant. Figure \ref{fig:pressuresignalexp} shows that the pressure signal varies from 2\,bar up to 4.9 bar during this experiment.

Figure\,\ref{fig:Druck2bar} shows all estimated signals. 
All signals in this figure are filtered by a first order low pass filter with $0.5$\,s filter time constant for ease of comparison.  
It is evident that the proposed algorithm estimates speed and torque with a high accuracy.  
The relative errors with respect to measured variables in steady state remain within ${0.05}$\% for the speed estimation,  $4$\% for the torque estimation, and $0.02$\% for the estimation of the system output $y$ defined in \eqref{eq:statespace}.
As apparent from the lower left diagram in Fig.\,\ref{fig:Druck2bar}, altering $\overline{p_D}$ leads to minor variations in $\tfrac{1}{2} \hat \theta_{p_D}-\theta_p$. This implies that $\theta_\text{off}$ is not perfectly constant, but weakly depends on $\overline{p_D}$. 
The lower right diagram in Fig.\,\ref{fig:Druck2bar} shows this dependence has a negligible effect on the estimation. 

\section{Conclusion}
We %
proposed an  algorithm for the speed, rotor position, and torque estimation of a motor-pump system including a progressive cavity pump. 
The algorithm is suited for applications in which the motor operates under V/f open-loop control. 
It only requires measurements of the pump discharge pressure and the effective value of the motor stator current and therefore allows to estimate the shaft torque and rotational speed without any expensive sensors. 
It is simple to implement the proposed method, and it shows a high estimation accuracy in our practical implementation. 
\section*{Acknowledgment}
The authors gratefully acknowledge the Ministerium f\"ur Wirtschaft, Innovation, Digitalisierung und Energie des Landes Nordrhein-Westfalen for funding this project.
\begin{figure}[H]
\centering
 \includegraphics[scale = .4]{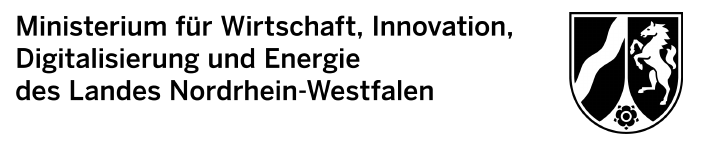}
\vspace*{.3cm}
 \end{figure}
 \section*{Appendix A: convergence of the extended Kalman filter based observer}
We how that $V(e) = e^\top P^{-1} e$ with $e = x -\hat x$ is a Lyapunov function for the observer \eqref{eq:observersyst}. 
Since $P(t)$ satisfies the Riccati equation \eqref{eq:schazer}, $P(t)^{-1}$ exits, is symmetric and thus its eigenvalues are real. As the eigenvalues of  $P(t)^{-1}$ are the inverse of the eigenvalues of $P(t)$, the matrix $P(t)^{-1}$ is positive definite if and only if $P(t)$ is finite and positive definite. We use the approach presented in \cite{Dieci1994} to prove positive definiteness of $P(t)$. First, consider the Lyapunov equation 
\begin{align}\label{eq:lyapeq}
\dot P(t) &= A(t) P(t) + P(t)A(t)^\top+Q. 
\end{align}
The solution to \eqref{eq:lyapeq} is given by
\begin{align}
\hspm P(t) = \phi(t,0) P(0)\phi(t,0)^\top +\int_{0}^t \phi(t,\tau) Q \phi(t,\tau)^\top d\tau,
\end{align}
where $\phi$ is the solution of 
\begin{align}
 \frac{\partial \phi(t,\tau)}{\partial t} = A(t) \phi(t,\tau), \quad \phi(\tau,\tau) = I, ~\text{for~all}~ t\geq \tau.
\end{align}
It follows from the fact $P(0)>0$ and non-singularity of $\phi(t,0)$ that $P(t)$ is positive definite. Suppose $Y(t)= \frac{1}{2}P(t)$. Then, 
\eqref{eq:schazer} can be re-written as
\begin{align} \label{eq:ricattieq}
 \dot P(t) & = [A(t)-Y(t)C(t)R^{-1}C(t)]^\top P(t)+\notag \\
& \hspace{1cm} +P(t)[A(t)-Y(t)C(t)R^{-1}C(t)] +Q,
\end{align}
which has the form of Lyapunov equation \eqref{eq:lyapeq}.
The solution of \eqref{eq:ricattieq} is non-negative as long as it exists. Since for symmetric non-negative matrices the condition $\|P\| =\text{max}_{\|\xi\|=1 } \xi^\top P \xi$ is valid, we conclude from the Riccati equation \eqref{eq:schazer} that
\begin{align}
 P(t)& = P(0) + \\
 &\hspace{-.75cm} +\hspace{-3pt} \int_0^t \hspace{-2pt} [A(\tau) P(\tau)\hspace{-2pt} + \hspace{-2pt} P(\tau)A(\tau)^\top\hspace{-3pt}-\hspace{-3pt} P(\tau)C^\top \hspace{-1pt} R^{-1}C(\tau)P(\tau)\hspace{-2pt}+\hspace{-2pt}Q] d\tau \notag,
\end{align}
and respectively the following inequality hold
\begin{align}
 \| P(t)\|\leq \|P(0)\| + \int_0^t [2 \|A(\tau)\|.\|P(\tau)\|+\|Q\|d\tau].
\end{align}
Hence, $\|P(t)\|$ is finite and $P(t)$ exists for all $t>0$. 
 Next, we investigate under which condition the derivative of $V$ is negative definite. For the computation of $\dot V(e)$, we first need to compute $\dot e$. Defining the vector 
 $ D = [x_1 ~ x_2 ~0 ~0]$, we can write
\begin{align}
\dot {{e}}(t) &= A(t) e(t) - K(t)(y(t) -\hat y(t))\notag\\
&= A(t)e(t) -\tfrac{1}{2}K(x_1^2 +x_2^2 -\hat x_1^2 - \hat x_2^2)\notag\\
& = A(t)e(t)- \tfrac{1}{2} K[x_1+\hat x_1~~ x_2+\hat x_2]\begin{bmatrix}  x_1- \hat x_1\\
                             x_2- \hat x_2
                    \end{bmatrix} \notag \\
&= A(t)e(t)-\tfrac{1}{2} K(t)\left(D+ C\right)e(t).
                   \end{align}
 Using $K(t)=P(t)C(t)^\top R^{-1}$ and $\dot P$ from \eqref{eq:schazer}, we compute $\dot V(e)$ as 
\begin{align}
 \dot V(e)&= - e^\top P^{-1} \dot P P^{-1}e+ 2 e^\top P^{-1}\dot e = \notag\\
 & = -e^\top P^{-1}(A^\top P +PA-PC^\top R^{-1}CP+Q)P^{-1}e +\notag\\
 & \hspace{.4cm}+2e^\top P^{-1} A e - e^\top C^\top R^{-1} D\, e-e^\top  C^\top R^{-1} C e\notag\\
 & = -e^\top P^{-1}QP^{-1}e - e^\top C^\top R^{-1} D\, e. 
 \end{align}
With $R= \kappa I$ for a positive number $\kappa$ and $Q>0$, the derivative of the Lyapunov function  is obviously negative definite if
$C^\top D\geq 0$, or equivalently, if 
\begin{align}
 [\hat x_1~~ \hat x_2]\begin{bmatrix}
                   x_1\\
                   x_2
                  \end{bmatrix} \geq 0. 
\end{align}
This inequality holds when the angle between the two vectors $[x_1~~ x_2]^\top$ and $[\hat x_1~~ \hat x_2]^\top$ is less than or equal to $\pi/2$. This condition is satisfied if $\text {sign}(x_1)=\text {sign}(\hat x_1)$ and $\text {sign}(x_2)=\text {sign}(\hat x_2)$. In other words, $\dot V(e)$ is negative definite as long as the estimated variables possess the correct signs. Under this condition the convergence of the observer is established. Note that the phase of the voltage vector of an induction motor always precedes the phase of the current vector and the angle between the voltage and current vectors is always smaller than $90$ degrees. This implies that  in the voltage coordinate system $i_{sd}$ is positive and $i_{sq}$ is negative.%
 \bibliographystyle{plain}
 \bibliography{ifacconf}
\end{document}